\documentclass[usenatbib]{mn2e}
\usepackage{apjfonts,amsfonts,amsmath,amssymb,bm,ctable,verbatim}

\newcommand{\gizmourl}{\href{http://www.tapir.caltech.edu/~phopkins/Site/GIZMO.html}{\url{http://www.tapir.caltech.edu/~phopkins/Site/GIZMO.html}}}

\newcommand{\etal}{et al.}

\newcommand{\acknowledgments}[1]{\begin{small}\section*{Acknowledgments}\end{small}{\noindent #1}\vspace{5pt}}
\newcommand{\datastatement}[1]{\begin{small}\section*{Data Availability Statement}\end{small}{\noindent #1}\vspace{5pt}}

\title[Why Bulges?]{Why do Black Holes Trace Bulges (\&\ Central Surface Densities), Instead of Galaxies as a Whole?}

\author[Hopkins \etal]{
\parbox[t]{\textwidth}{
Philip F.~Hopkins$^1$, Sarah Wellons$^2$, Daniel Angl{\'e}s-Alc{\'a}zar$^3$, \\
Claude-Andr{\'e} Faucher-Gigu{\`e}re$^2$, \&\ Michael Y.\ Grudi{\'c}$^2$
}\vspace*{4pt} \\
$^1$ TAPIR, Mailcode 350-17, California Institute of Technology, Pasadena, CA 91125, USA. E-mail:phopkins@caltech.edu \\
$^2$ CIERA and Department of Physics and Astronomy, Northwestern University, 2145 Sheridan Road, Evanston, IL 60208, USA \\
$^3$ Department of Physics, University of Connecticut, 196 Auditorium Road, U-3046, Storrs, CT 06269-3046, US}

\date{}
\begin{document}
\maketitle

\begin{abstract}
Previous studies of fueling black holes (BHs) in galactic nuclei have argued (on scales $\sim0.01-1000\,$pc) accretion is dynamical with inflow rates  $\dot{M}\sim\eta\,M_{\rm gas}/t_{\rm dyn}$ in terms of gas mass $M_{\rm gas}$, dynamical time $t_{\rm dyn}$, and some $\eta$. But these models generally neglected expulsion of gas by stellar feedback, or considered extremely high densities where expulsion is inefficient. Studies of star formation, however, have shown on sub-kpc scales the expulsion efficiency $f_{\rm wind}=M_{\rm ejected}/M_{\rm total}$ scales with the gravitational acceleration as $(1-f_{\rm wind})/f_{\rm wind}\sim\bar{a}_{\rm grav}/\langle \dot{p}/m_{\ast}\rangle \sim\Sigma_{\rm eff}/\Sigma_{\rm crit}$ where $\bar{a}_{\rm grav}\equiv G\,M_{\rm tot}(<r)/r^{2}$ and  $\langle\dot{p}/m_{\ast}\rangle$ is the momentum injection rate from young stars. Adopting this as the simplest correction for stellar feedback, $\eta\rightarrow\eta\,(1-f_{\rm wind})$, we show this provides a more accurate description of simulations with stellar feedback at low densities. This has immediate consequences, predicting the slope and normalization of the $M_{\rm BH}-\sigma$ and $M_{\rm BH}-M_{\rm bulge}$ relation, $L_{\rm AGN}-$SFR relations, and explanations for outliers in compact Es. Most strikingly, because star formation simulations show expulsion is efficient ($f_{\rm wind}\sim1$) below total-mass surface density $M_{\rm tot}/\pi\,r^{2}<\Sigma_{\rm crit}\sim3\times10^{9}\,M_{\odot}\,{\rm kpc^{-2}}$ (where $\Sigma_{\rm crit}=\langle\dot{p}/m_{\ast}\rangle/(\pi\,G)$), BH mass is predicted to specifically trace host galaxy properties above a critical surface brightness $\Sigma_{\rm crit}$ (B-band $\mu_{\rm B}^{\rm crit}\sim19\,{\rm mag\,arcsec^{-2}}$). This naturally explains why BH masses preferentially reflect bulge properties or central surface-densities (e.g.\ $\Sigma_{1\,{\rm kpc}}$), not `total' galaxy properties. 
\end{abstract}


\section{Introduction}
\label{sec:intro}

Understanding the origins, growth and evolution of super-massive black holes (BHs) remains one of the most important unsolved problems in extragalactic astrophysics. It is now well-established that most sufficiently-massive galaxies host BHs whose masses correlate with various host galaxy bulge properties (\citealt{magorrian,FM00,Gebhardt00,hopkins:bhfp.obs,aller:mbh.esph,kormendy:2011.bh.nodisk.corr}; for a review see \citealt{kormendy:2013.review.smbh.host.correlations}). The small scatter in these correlations \citep[relative to other galaxy properties;][]{hopkins:msigma.scatter}, together with constraints indicating that most BH mass is assembled in an optically bright quasar phase \citep{Soltan82,salucci:bhmf,yutremaine:bhmf,hopkins:old.age}, suggests a picture of ``co-evolution'' between galaxies and accreting BHs visible as active galactic nuclei (AGN) or quasars \citep{merloni:synthesis.model}. Understanding this ``co-evolution'' has far-reaching consequences beyond the black holes themselves: for example, it is widely believed that ``feedback'' from accreting BHs (in the form of radiation, winds, and jets; \citealt{laor:warm.absorber,crenshaw:nlr,dunn:agn.fb.from.strong.outflows,sturm:2011.ulirg.herschel.outflows,Zakamska2016,Williams2017}) can unbind, expel, or super-heat gas in the vicinity of the BH and throughout the host galaxy \citep{silkrees:msigma,king:msigma.superfb.1,dimatteo:msigma,murray:momentum.winds,hopkins:lifetimes.methods,hopkins:lifetimes.obscuration,debuhr:momentum.feedback,torrey:2020.agn.wind.bal.gal.fx.fire}, potentially regulating star formation and galaxy stellar masses \citep{croton:sam,hopkins:qso.all,hopkins:groups.ell} and the structure of the circum-galactic medium around massive galaxies \citep{ciottiostriker:cooling.flow.selfreg.1,cox:xray.gas,best:radio.loudness,Voit_2017}. 

But modeling the strength of ``feedback'' from SMBHs, and their presence in the first place, depends fundamentally on understanding their accretion rates. In understanding how gas is transported from the inter-galactic medium onto black holes, it is especially important to understand, both empirically and theoretically, how gas is transported from scales $\sim 0.1-1000\,$pc within the galaxy (where its angular momentum is $\sim 10^{7}$ times too large to be accreted by the BH directly) into the BH accretion disk (scales $\lesssim 0.01\,$pc). These scales include the observational and numerical resolution limits of essentially all resolved galaxy surveys and/or galaxy-scale numerical simulations \citep{fabian:2012.agn.fb.obs.review,schartmann:2010.1068.star.cluster.fueling,hopkins:2013.fire,2017ARA&A..55...59N,2019MNRAS.486.2827D} -- so in both empirical and theoretical studies of AGN ``fueling'' and its relation to galaxy properties, these are the key scales one wishes to relate to the AGN accretion rate. Moreover, neither well-understood galaxy-scale angular momentum transport mechanisms (mergers, galaxy-scale arms/bars), nor well-understood traditional accretion-disk processes (e.g.\ the MRI and turbulent/viscous stresses), can operate efficiently over most of these scales (especially from $\sim 0.01 - 10\,$pc, within the BH radius of influence), leading to one of several ``last parsec problems'' \citep{goodman:qso.disk.selfgrav,jiang:low-res.leads.to.underestimated.fragmentation}. Moreover the assumptions of the classical Bondi-Hoyle \citep{1944MNRAS.104..273B} or \citet{shakurasunyaev73} type accretion models are violated by many orders of magnitude on these scales: gas within a galaxy is rapidly-cooling ($t_{\rm cool} \ll t_{\rm freefall}$), self-gravitating, star-forming, turbulent, must lose most of its angular momentum to efficient torques to be accreted, and the potential is dominated by a combination of gas, collisionless stars and dark matter, and the black hole itself \citep{hopkins:zoom.sims,hopkins:inflow.analytics,daa:20.hyperrefinement.bh.growth}.

Empirically, it is clear that the best galactic {\em predictors} of BH mass on these scales are the velocity dispersion and/or stellar mass of the central classical ``bulge,'' or nuclear star cluster (NSC) in late-type dwarf galaxies which exhibit no classical bulge, as opposed to e.g.\ {\em total} galaxy stellar or disk or halo mass or luminosity or circular velocity \citep[e.g.][]{mancini:2012.bh.mass.vs.KE.gal,kormendy:2013.review.smbh.host.correlations,reines:2015.dwarf.gal.mbh.mgal.norm.dift.and.huge.scatter}. But this itself presents an important theoretical puzzle, related to the question above of what physics actually drives accretion on these scales. Almost all theoretical models to date of BH mass growth via pure accretion (i.e.\ ``fueling-limited'' models), hierarchical assembly (e.g.\ BH growth primarily via mergers), and/or self-regulation via feedback (i.e.\ ``feedback-regulated'' models) predict correlations between BH mass and ``gas supply in the galaxy center'' or ``depth of the potential'' in which the BH sits or ``mass assembled via mergers'' (e.g.\  \citealt{silkrees:msigma,king:msigma.superfb.1,dimatteo:msigma,hopkins:bhfp.theory,peng:2007.mbh.mhost.from.central.limit.theorem}). These models commonly assume that these properties correlate closely with ``bulge'' or NSC mass, but that is not correct in galaxies that are not  bulge-dominated. 

For example, in almost all galaxies of Sa or later type (including the Milky Way), the bulge {\em does not} dominate the central potential, relative to either the stellar+gas disk/entire galaxy or the dark matter: this can be seen from simple comparison of $G\,M_{\rm bulge}/R_{\rm bulge}$ vs $G\,M_{\rm disk}/R_{\rm disk}$ and $G\,M_{\rm halo}/R_{\rm halo}$, or more detailed Jeans modeling \citep{aller:mbh.esph,2017MNRAS.465...76M,2017ApJ...850...70T}. The discrepancy can be orders-of-magnitude in dwarfs.\footnote{Using standard abundance-matching relations from \citet{behroozi:2019.sham.update} and assuming \citet{nfw:profile} halos, the central potential from the DM {\em alone} in sub-$L_{\ast}$ (dwarf) galaxies scales as $\Phi(r\rightarrow0) \sim (250\,{\rm km\,s^{-1}}\,[M_{\ast}/10^{10}\,M_{\odot}]^{1/6})^{2}$ 
%
-- much larger than the potential from the bulge or NSC (or stellar disk), and very weakly dependent on stellar mass, while e.g.\ the BH and bulge/NSC mass scales super-linearly with stellar mass as $M_{\rm BH} \propto M_{\rm bulge/NSC} \propto M_{\ast}^{2-4}$ \citep{reines:2015.dwarf.gal.mbh.mgal.norm.dift.and.huge.scatter,graham:2015.mbh.host.steep.at.lowmass}.} 
However, if one considers the potential gradients, i.e.\ gravitational {\em acceleration} provided by these components ($\propto G\,M/R^{2}$), then the bulge often does dominate in the region between the SMBH radius of influence and the outskirts of the bulge -- a crucial difference to which we will return below.

The ``gas supply to the galaxy center'' is also not particularly well-correlated with the bulge mass: nuclear bulge/cluster/disk shapes/densities/masses/radii vary wildly \citep{ferrarese:type12,lauer:bimodal.profiles,lauer:central.minimum.ell,2016ApJS..222...10S,2016ApJ...825....3L}. So there is no reason, in most models for BH growth, why BHs would correlate particularly well with the ``central mass'' within an arbitrarily varying annulus $<R$ that happens to correspond to the ``bulge'' size. 
Yet \citet{hopkins:msigma.scatter} argued that the total bulge/NSC mass was, in practice, often a better predictor of $M_{\rm BH}$ compared to e.g.\ mass within a fixed physical annulus or multiple of the BH ``radius of influence'' $R_{\rm ROI} \sim G\,M_{\rm BH}/\sigma^{2}$.
And in dwarfs, the observed SMBHs/AGN and their associated light excess/``bulge'' are in fact  most often {\em not} located near the center-of-mass or center-of-light of the galaxy \citep{reines:2020.off.nuclear.agn.in.dwarfs}, if such a center can even be defined (it often cannot at $\lesssim\,$kpc scales). 
Regarding ``mass assembled by mergers,'' it is increasingly clear that in sub-$L_{\ast}$ galaxies galaxy-galaxy mergers play a minor/secondary role in bulge formation \citep{1996ApJ...457L..73C,governato:2010.dwarf.gal.form,hopkins:merger.rates,hopkins:merger.rates.methods,puech:2012.disk.survival,2015ApJ...799..184P}; even if they do, most of the ``incoming'' mass associated with such mergers ends up in an extended halo, rather than a compact bulge, and produces relatively little contribution to BH growth/AGN activity in dwarfs, $\sim L_{\ast}$, or Seyfert galaxies \citep{2012ApJ...744..148K,2012NewAR..56...93A,hopkins:2013.agn.host.morph,2014ARA&A..52..589H}. Finally, {\em none} of the theoretical models described above explain why BHs would correlate more poorly with ``pseudo-bulges'' and ``nuclear disks'' as defined photometrically following \citet{kormendy.kennicutt:pseudobulge.review}, as compared to ``classical'' photometric bulges.

In this letter, we combine qualitative scalings common to many of the accretion models described above with a simple correction, generally neglected in simulation prescriptions, for the mass fraction expelled by {\em stellar} feedback from star formation on sub-kpc scales (``between'' the simulation-resolved scales and accretion disk), and show that this provides an immediate and natural resolution to the questions above.

\section{Theory}
\label{sec:theory}

The problem of accretion from sub-kpc scales described in \S~\ref{sec:intro} has been studied in detail in many papers, for example the series by \citet{hopkins:zoom.sims,hopkins:cusp.slopes,hopkins:inflow.analytics,hopkins:m31.disk}, subsequently explored further in other work \citep[e.g.][and others discussed below]{hopkins:torus,angles-alcazar:2013.bh.growth.vs.accretion.prescription,angles.alcazar:grav.torque.accretion.cosmo.sim.implications,daa:BHs.on.FIRE,2019MNRAS.486.2827D,thomas:bhs.in.simba.sims,daa:20.hyperrefinement.bh.growth}. These studies generically showed that on these scales, accretion is regulated by ``gravitational torques'' from a combination of asymmetries in the potential, interactions between the collisionless (stars+dark matter) and gas components, and shocks/dissipation in the gas, giving rise to an accretion rate of the form:
\begin{align}
\label{eqn:mdot.basic} \dot{M}_{\rm acc} &= \eta\,M_{\rm gas}\,\Omega \sim \eta\,\frac{f_{\rm gas}\,V_{c}^{3}}{G} \sim \eta\,\frac{4\pi\,G^{2}M_{\rm tot}^{2}\rho}{V_{c}^{3}} 
\end{align}
where $M_{\rm gas}\approx \pi\,\Sigma_{\rm gas}\,R^{2}$ is the gas mass within some annulus $R$, $\Omega = V_{c}/R$ is the dynamical frequency, and $\eta$ is some relatively-weakly-varying function which describes the magnitude of whatever torques actually remove angular momentum and allow for accretion. For example, in the model from \citet{hopkins:inflow.analytics} $\eta \approx 0.01\,(M_{\rm S}/M_{\rm d})^{1/6} \, [1 + 3\,M_{\rm d,\,9}^{1/3}\,(M_{\rm gas}/M_{\rm d})]^{-1} \sim 0.001$ where $M_{\rm S} = M_{\rm BH} + M_{\rm \alpha\,disk}$ is the total ``sink'' (BH+accretion disk) system mass, and $M_{\rm d,\,9}\equiv M_{\rm d}/10^{9}\,M_{\odot}$ with $M_{\rm d}(<R)$ the total mass in a ``disky'' (rotation-dominated) component. 
A number of subsequent, independent idealized theoretical studies \citep{2012ApJ...758...14K,2013ApJ...771..119A,2015ApJ...806..150L,emsellem:nuclear.fueling.sims,2019MNRAS.486.5377I} have validated the qualitative scaling above for similar assumptions, and detailed observations of galactic nuclei have appeared to confirm both the dominance of gravitational torques, and the approximate scaling of inflow rates with dynamical nuclear properties as predicted by these models \citep{2013A&A...558A.124C,2014A&A...567A.125G,esquej:torque.model.nuclear.sf.vs.mdotbh,querejeta:grav.torque.obs.m51}. Broadly speaking, even quite different accretion models have arrived at scalings which qualitatively follow Eq.~\ref{eqn:mdot.basic} on similar scales.\footnote{For example, (1) assuming a constant accretion rate per free-fall time simply gives $\eta=$\,constant, by definition. (2) The ``gravito-turbulent''-type models motivated by \citet{gammie:2001.cooling.in.keplerian.disks}, applied to star-forming disks with Toomre $Q\sim 1$ as in \citet{thompson:rad.pressure,kawakatu:disk.bhar.model,hopkins:2013.turb.planet.direct.collapse} give $\eta\approx 0.1\,(M_{\rm d}/M_{\rm tot})^{2} \sim $\,constant. (3) ``Ballistic accretion'' \citep{hobbs:turbulence.agn.feeding} gives $\eta \approx (h/R)^{-1}\,\exp{(-0.6\,R^{2}/h^{2})}$ which is constant if the disks are thick ($h\sim R$) or we assume $h/R\sim\,$constant, or scales similarly to ``gravitoturbulent'' cases if we take $Q\sim$\,constant. (4) A generalized version of the \citet{shu:isothermal.sphere.collapse} self-similar scaling for a collapsing isothermal sphere, allowing for non-gas contributions to the potential and turbulence, gives $\eta \approx (1+\Delta_{v}^{2})^{-3/2}$ with $\Delta_{v}^{2} \equiv (c_{s}^{2} + \sigma_{\rm turb}^{2}/3 + |\langle \delta {\bf v} \rangle|^{2})/V_{c}^{2}$ (with sound speed $c_{s}$, 3D gas velocity dispersion $\sigma_{\rm turb}$, and bulk BH-gas relative velocity $\delta {\bf v}$; see \citealt{hopkins:qso.all,dimatteo:cosmo.bhs}). (5) The estimator in e.g.\ \citet{hobbs:2012.freefall.agn.prescription} for ``Bondi-like'' accretion in a halo (ignoring turbulence and relative motion) is simply this with $\eta = (1+c_{s}^{2}/V_{c}^{2})^{-3/2} \approx 1$.}

However, all of these studies essentially neglected the possibility that gas would be efficiently expelled from the galactic nucleus by {\em stellar} feedback (e.g.\ radiation pressure, stellar mass-loss, and SNe explosions), before it could accrete into the BH accretion disk. This includes models which treat stellar feedback as a ``sub-grid'' process influencing the ISM but either not driving strong outflows or simply driving outflows with a by-hand fixed ``efficiency'' $\dot{M}_{\rm out} \sim \dot{M}_{\ast}$, as well as those which neglect it entirely. A couple of subsequent studies \citep[e.g.][]{hopkins:qso.stellar.fb.together,wada:torus.mol.gas.hydro.sims,kawakatu:2020.obscuration.torus.from.stellar.fb.in.torus,daa:20.hyperrefinement.bh.growth} have revisited this problem with simulations that explicitly include the relevant stellar feedback processes. However, these were simulations of nuclear disks intended to model extremely bright QSOs with enormous surface mass densities (or accelerations), $\Sigma_{\rm eff} \equiv M_{\rm tot}(<R) / \pi\,R^{2} \gtrsim 10^{5}\,{\rm M_{\odot}\,pc^{-2}}$, where stellar feedback (even from vigorous SNe explosions) is unable to unbind large quantities of gas, and served primarily to ``thicken'' the nuclear disk (potentially explaining features of the obscuring ``torus''; \citealt{wada:starburst.torus.model,thompson:rad.pressure}).

Under less-extreme conditions, many theoretical \citep{wutschik:2013.model.sf.near.gal.nuclei,torrey.2016:fire.galactic.nuclei.star.formation.instability,grudic:sfe.cluster.form.surface.density,grudic:max.surface.density} and observational \citep{vollmer:2008.torus.evolution.stellar.fb,izumi:nuclear.disk.plus.outflow.equals.mass.acc} studies have pointed out that stellar feedback can in principle easily expel most of the gas from galactic nuclei, dramatically suppressing accretion rates onto the BH.  
This can occur ``indirectly'' or ``directly.'' In the  ``indirect'' sense, efficient stellar feedback can, in a cosmological sense, lead to a given dark matter halo producing a much-less-massive,  lower-density galaxy, which in turn produces sub-kpc conditions less conducive to BH growth \citep[see discussion in e.g.][]{bower:2017.bh.growth.quenching.via.stellar.agn.interplay,habouzit:2017.bh.growth.cosmosims.suppressed.by.sne.fb}. These effects would therefore be implicit in the accretion models discussed above. But stellar feedback can also ``directly'' restrict accretion through a given annulus in a galaxy given fixed larger-scale conditions, by ejecting some of that material in the annulus which would otherwise have lost its angular momentum \citep[e.g.][]{dubois:delayed.cooling.sne.models,grudic:max.surface.density}. The latter is the case of interest here. While such behavior has been qualitatively observed in simulations, a simple quantitative parameterization of its effects is still lacking. 
Therefore consider: the simplest parameterization of this effect is to take 
\begin{align}
\eta &\rightarrow \left( 1-f_{\rm wind} \right) \, \eta 
\end{align}
where $f_{\rm wind} \equiv M_{\rm ejected} / M_{\rm gas,\,total}$ represents the fraction of gas expelled by stellar feedback from within the annulus. 

As shown in detail in \citet{torrey.2016:fire.galactic.nuclei.star.formation.instability}, in galactic nuclei, the scalings for star formation and $f_{\rm wind}$ are essentially the same as in massive GMC complexes, as opposed to ``galactic'' outflow/star formation models. This is fundamentally because on spatial scales $\sim 0.1-1000\,$pc, the dynamical times $t_{\rm dyn} \sim 0.5\,{\rm Myr}\,(R/100\,{\rm pc})\,(200\,{\rm km\,s^{-1}}/V_{c})$ are much shorter than the $t_{\rm fb} \sim 30-100\,$Myr timescales over which most stellar feedback is deposited. So gas flows in, converts to stars on some number of free-fall times (as in a ``single burst''), but the stars formed then rapidly expel gas from the central regions as they age and SNe begin to explode (akin to GMC destruction): no ``steady-state'' is possible when $t_{\rm dyn} \ll t_{\rm fb}$. 

A simple analytic model for $f_{\rm wind}$ in this limit is given by \citet{fall:2010.sf.eff.vs.surfacedensity}, as updated in \citet{grudic:mond.accel.scale.from.stellar.fb}: upon forming, a mass $M_{\rm \ast,\,young}$ of young stars ($\ll 100\,$Myr old) within the nucleus in an area $A \sim \pi\,R^{2}$ will inject momentum into the surrounding gas (via feedback) at a rate: 
\begin{align}
\label{eqn:pfb} \frac{d \dot{P}_{\rm fb}}{d A} \sim \langle \dot{p}/m_{\ast} \rangle\, \frac {M_{\rm \ast,\, young}}{A}
\end{align}
where 
\begin{align}
\langle \dot{p}/m_{\ast} \rangle \sim ({\rm a\ few})\,\frac{L_{\ast}/c}{m_{\ast}} \sim 1000\,\frac{L_{\odot}}{M_{\odot}\,c} \sim 10^{-7}\,\frac{\rm cm}{\rm s^{2}}
\end{align}
is the momentum injection rate per stellar mass, for a well-sampled IMF.\footnote{Crucially, the quantity $\langle \dot{p}/m_{\ast} \rangle$ for a ``young'' (ZAMS or age $\lesssim 30\,$Myr) is approximately independent of whether the dominant stellar feedback comes from radiation pressure, expanding HII regions, O/B winds, or SNe; see \citet{starburst99,BC03,hopkins:fb.ism.prop,agertz:2013.new.stellar.fb.model,2015ApJ...815...67K}.} This will expel the remaining gas when $d \dot{P}_{\rm fb}/{d A}$ exceeds the force per unit area on the gas from gravity:
\begin{align}
\label{eqn:fgrav} \frac{d {\rm Force}_{\rm grav}}{d A} \sim \bar{a}_{\rm grav}\,\frac{M_{\rm gas}}{A} \sim \frac{G\,M_{\rm tot}}{R^{2}}\,\frac{M_{\rm gas}}{R^{2}} \sim G\,\Sigma_{\rm eff}\,\Sigma_{\rm gas}\ , 
\end{align}
with 
\begin{align}
\bar{a}_{\rm grav} &\equiv \frac{G\,M_{\rm tot}(<r)}{r^{2}} \\ 
\Sigma_{\rm eff} &\equiv \frac{M_{\rm tot}(<r)}{\pi\,r^{2}}
\end{align}
defined inside a spherical annulus of radius $r$. Equating $d {\rm Force}_{\rm grav}/dA$ (Eq.~\ref{eqn:fgrav}) and ${d \dot{P}_{\rm fb}}/{d A}$ (Eq.~\ref{eqn:pfb}) and solving for $M_{\rm gas}$ to obtain the gas mass which can be expelled gives:
\begin{align}
\label{eqn:mexpelled} \frac{M_{\rm gas,\,expelled}}{M_{\rm \ast,\, young}} \sim \frac{M_{\rm ejected}}{M_{\rm retained}} = \frac{f_{\rm wind}}{1-f_{\rm wind}} \sim \frac{\langle \dot{p}/m_{\ast} \rangle}{\bar{a}_{\rm grav}} \ , 
\end{align}
i.e.\ 
\begin{align}
\label{eqn:fwind}
1-f_{\rm wind} &\approx \frac{\bar{a}_{\rm grav}}{\langle \dot{p}/m_{\ast} \rangle + \bar{a}_{\rm grav}} = \frac{\Sigma_{\rm eff}}{\Sigma_{\rm crit} + \Sigma_{\rm eff}} 
\end{align}
with 
\begin{align}
\Sigma_{\rm crit} = \frac{\langle \dot{p}/m_{\ast} \rangle}{\pi\,G} \sim 3000\,\frac{\rm M_{\odot}}{\rm pc^{2}} = 3\times10^{9}\,\frac{\rm M_{\odot}}{\rm kpc^{2}} \sim 0.6\,\frac{\rm g}{\rm cm^{2}}\ .
\end{align}
Because, in essentially all reasonable models on scales $\sim 1-1000\,$pc, {\em most} of the retained mass goes into star formation rather than inflow to the BH, we can safely neglect the correction for inflow itself in this derivation of $f_{\rm wind}$.\footnote{More formally, there is some subtle ambiguity in Eq.~\ref{eqn:mexpelled} in how precisely to relate $M_{\rm retained}$, $M_{\rm \ast,\,young}$, and some (generally much smaller) mass accreted through an annulus in the same time, which can be addressed more accurately with the continuum limit models discussed below. But since this is simply an order-of-magnitude argument and the behavior is identical in the relevant limits we study below, we can neglect this ambiguity for now.}

Alternatively, adopting a continuum limit within each annulus as gas moves to the BH, we can revisit the derivation of Eq.~\ref{eqn:mdot.basic} in \citet{hopkins:inflow.analytics}. There, we solved a steady-state model calculating the strength of gravitational torques driving a {\em total} inflow rate within each annulus, coupled to the continuity equation, with $\dot{M}_{\rm in}(R) = \dot{M}_{\rm in}(R+dR) - \dot{M}_{\ast}(R < R^{\prime} < R+dR) = \dot{M}_{\rm in}(R+dR) - 2\pi\,R\,dR\,\dot{\Sigma}_{\ast}(R)$, i.e.\ accounting for gas lost to star formation within each annulus. If we modify this to also include gas lost in winds, then $\dot{\Sigma}_{\ast}(R) \rightarrow \dot{\Sigma}_{\ast} + \dot{\Sigma}_{\rm wind} = (1+\eta_{\rm wind}[R])\,\dot{\Sigma}_{\ast}$ where $\eta_{\rm wind} \equiv \dot{\Sigma}_{\rm wind}/\dot{\Sigma}_{\ast} \sim \langle \dot{p}/m_{\ast} \rangle / \bar{a}_{\rm grav}$ within each annulus. While the exact solutions to this are, in general, numerical, simply taking $\eta\rightarrow \eta\,(1-f_{\rm wind})$ with $f_{\rm wind}$ from Eq.~\ref{eqn:fwind} provides a remarkably good approximation to the full solution, and is exact in small and large $\Sigma_{\rm eff}(R)$ limits. Since $\Sigma_{\rm eff}$ in the analytic model increases monotonically as $R\rightarrow 0$, the ``loss'' term $f_{\rm wind}$ is dominated by the largest radii, e.g.\ $R$ where it is evaluated: for $\Sigma_{\rm eff} \gg \Sigma_{\rm crit}$, the exact solution is unmodified from \citet{hopkins:inflow.analytics}, for $\Sigma_{\rm eff} \ll \Sigma_{\rm crit}$, it is multiplied by one power of $\Sigma_{\rm eff}/\Sigma_{\rm crit}$, as expected.

\section{Comparison to Numerical Simulations}
\label{sec:compare.numerical}

\begin{figure}
    \centering
    \includegraphics[width=0.97\columnwidth]{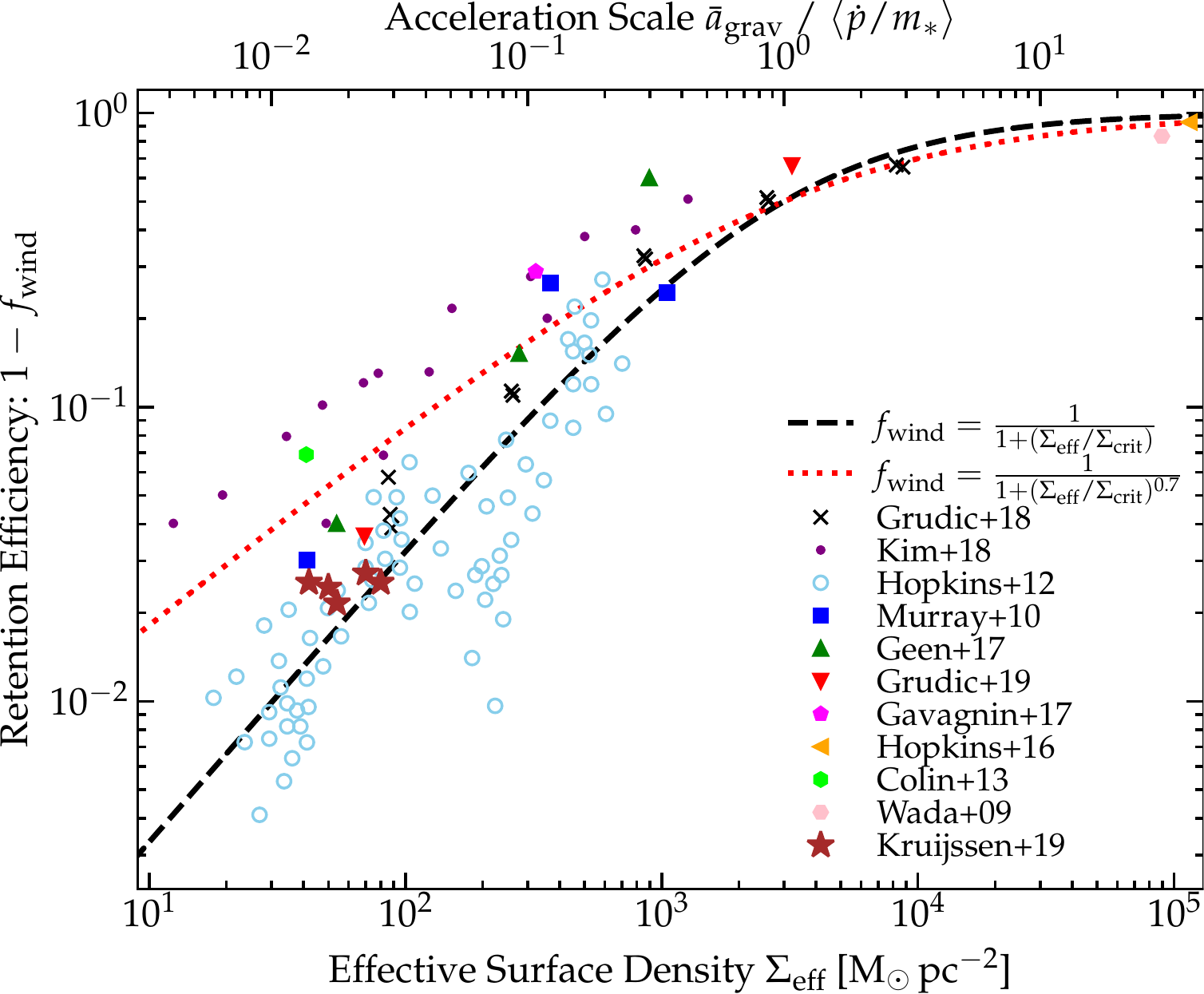}
    \caption{Scaling of the ``retention factor'' or wind loss factor $f_{\rm wind} \equiv M_{\rm ejected} / M_{\rm gas,\,total}$ measured in simulations and observations of $\lesssim 100\,$pc-scale structures: simulations of molecular clouds \citep{colin:2013.star.cluster.rhd.cloud.destruction,2017MNRAS.472.4155G,2017MNRAS.471.4844G,2018ApJ...859...68K,grudic:2019.imf.sampling.fx.on.gmc.destruction}, simulations including galactic nuclei \&\ disks \citep{hopkins:fb.ism.prop,grudic:sfe.cluster.form.surface.density}, circum-BH disk simulations \citep{wada:torus.mol.gas.hydro.sims,hopkins:qso.stellar.fb.together}, and observed GMCs \citep{2010ApJ...709..424M} and galactic nuclei \citep{2019Natur.569..519K}. We compare the simple predicted theoretical scaling from \citet{fall:2010.sf.eff.vs.surfacedensity,grudic:mond.accel.scale.from.stellar.fb} (Eq.~\ref{eqn:fwind}; $f_{\rm wind}^{-1} = 1+(\Sigma_{\rm eff}/\Sigma_{\rm crit}) = 1+ \bar{a}_{\rm grav} / \langle \dot{p}/m_{\ast} \rangle$), and a slight (arbitrary) variant fit ($f_{\rm wind}^{-1} = 1+(\Sigma_{\rm eff}/\Sigma_{\rm crit})^{0.7}$) which illustrates the theoretical uncertainties.
    \label{fig:fwind}}
\end{figure}

Eq.~\ref{eqn:fwind} is actually remarkably well-supported by both explicit numerical MHD simulations of GMC/star cluster/nuclear disk formation with explicit, resolved stellar feedback physics \citep{colin:2013.star.cluster.rhd.cloud.destruction,2017MNRAS.472.4155G,2017MNRAS.471.4844G,grudic:sfe.cluster.form.surface.density,grudic:2019.imf.sampling.fx.on.gmc.destruction,2018ApJ...859...68K} as well as observations \citep{vollmer:2008.torus.evolution.stellar.fb,2010ApJ...709..424M,grudic:sfe.gmcs.vs.obs,2019Natur.569..519K}, as shown in Fig.~\ref{fig:fwind}.\footnote{Since these are idealized simulations, $f_{\rm wind}$ can be easily measured as the fraction of  the initial gas mass which is entirely expelled.} A wide range of different numerical codes, methods, and treatments of stellar feedback, including simulations of both GMCs as well as nuclear stellar disks support such a scaling. We compare e.g.\ an arbitrary variant dependence on $\Sigma_{\rm eff}/\Sigma_{\rm crit}$ fitting different simulations that give somewhat different detailed behavior,\footnote{Note the difference between the somewhat-larger efficiency predicted by \citet{2018ApJ...859...68K} in Fig.~\ref{fig:fwind} and other plotted cases owes in part to the fact that \citet{2018ApJ...859...68K} included only UV radiation as a stellar feedback mechanism, but also to more detailed numerical and methodological differences discussed in detail therein and in \citet{hopkins:2019.grudic.photon.momentum.rad.pressure.coupling,grudic:2019.imf.sampling.fx.on.gmc.destruction}.} in order to illustrate that even with systematic differences in physics and numerical methods, different simulations predict a relation qualitatively similar to our simple order-off-magnitude estimate. These scalings and the results in Fig.~\ref{fig:fwind} also immediately explain why the previous simulations of ``QSO-scale nuclear disks'' discussed above, with $\Sigma_{\rm eff} \sim 10^{5}\,{\rm M_{\odot}\,pc^{-2}} \gg \Sigma_{\rm crit}$ (so $1-f_{\rm wind} \approx 1$) saw essentially negligible effects on the accretion rate scaling (compared to Eq.~\ref{eqn:mdot.basic}) including explicit stellar feedback, while lower-resolution cosmological simulations of high-redshift, lower-mass galaxies  (primarily dwarfs in low-luminosity AGN phases), with $\Sigma_{\rm eff} \ll \Sigma_{\rm crit}$ at their resolution limits, found that stellar feedback tended to ``blow out'' most of the gas ($1-f_{\rm wind} \ll 1$) before it could accrete, dramatically suppressing $\dot{M}_{\rm BH}$ \citep{dubois:delayed.cooling.sne.models,habouzit:2017.bh.growth.cosmosims.suppressed.by.sne.fb}.

\begin{figure}
    \centering
    \includegraphics[width=0.97\columnwidth]{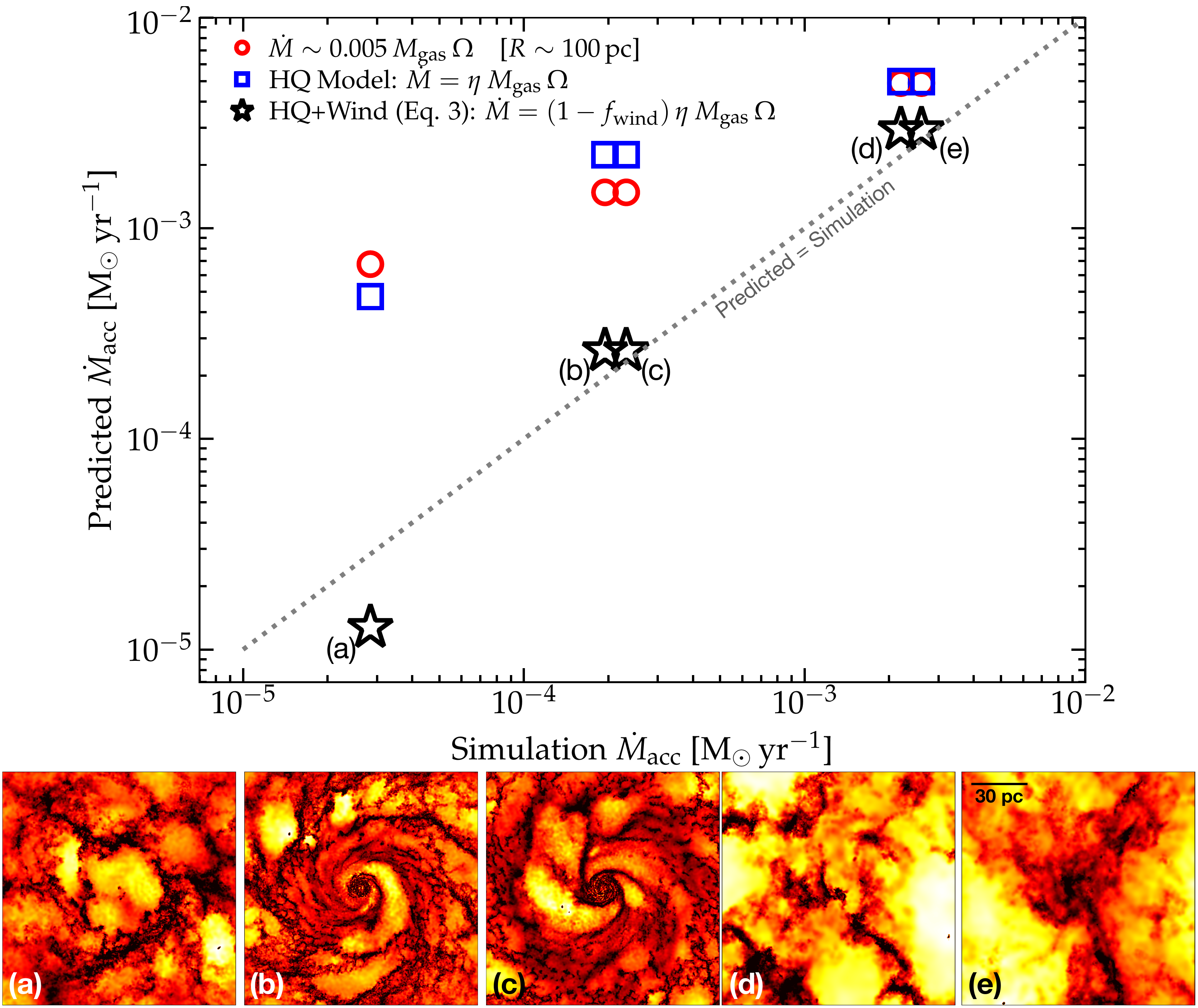}
    \caption{Direct test of different BHAR estimators in 5 simulations: we simulate a BH surrounded by an exponential gas+stellar disk (scale-length $100\,$pc) with star formation and stellar feedback as in \citet{hopkins:qso.stellar.fb.together}, with initial masses ($M_{\rm BH}$, $M_{\rm gas}$, $M_{\ast,\,{\rm disk}}$)$/M_{\odot}$ of (1e5,\,5e6,\,5e6) ({\bf a}), (1e7,\,5e6,\,5e6) ({\bf b}+{\bf c}), (1e7,\,5e8,\,5e8) ({\bf d}+{\bf e}), each run for $\sim 10$ dynamical times. Models {\bf b},{\bf c} and {\bf d},{\bf e} differ in the initial value of $Q=2.5,\,0.5$, respectively, given to the disk. We measure $\dot{M}_{\rm acc}$ as resolved gravitational capture of bound gas within $<0.1\,$pc (averaged over the simulation duration), and compare to the predicted $\dot{M}_{\rm acc}$ from the reference models, evaluating $M_{\rm gas}$, $\Omega$, $\Sigma_{\rm eff}$, etc.\ at $R=100\,$pc and $t=0$. We compare: $\eta=0.005=$\,constant; ``HQ,'' the \citealt{hopkins:inflow.analytics} model for $\eta$ in Eq.~\ref{eqn:mdot.basic}; and ``HQ+Wind,'' our proposed correction to this taking $f_{\rm wind}$ from Eq.~\ref{eqn:fwind}. Images show a gas density projection for each run (scale bar labeled in (e)); at the lowest $\Sigma_{\rm eff}$ ({\bf a}) we directly see stellar feedback evacuating the nuclear region.
    \label{fig:inflow.sims}}
\end{figure}

Fig.~\ref{fig:inflow.sims} tests this explicitly in high-resolution numerical simulations of inflow rates into the central $\ll 1\,$pc around a BH from $\sim 1-1000\,$pc radii disks, including detailed stellar feedback models identical to the no-AGN-feedback (``No\_BAL'')  simulations in \citet{hopkins:qso.stellar.fb.together}. We repeat their previous simulations with the same code and physics: the only difference is that we rescale the initial nuclear disk and BH masses such that $\Sigma_{\rm eff}$ ranges from $\sim 10^{2}-10^{4}\,M_{\odot}\,{\rm pc^{-2}}$, while they considered only a case with $\Sigma_{\rm eff} \sim 10^{5}\,M_{\odot}\,{\rm pc^{-2}}$. We compare the true accretion rates predicted by the high resolution sims, to the accretion rate that would be inferred by an analytic estimator based on the global simulation initial conditions or as a sub-grid model in a simulation at a lower, more typical resolution.  As expected, assuming $f_{\rm wind} \approx 0$ (i.e.\ taking Eq.~\ref{eqn:mdot.basic} without modification) works increasingly well at the highest $\Sigma_{\rm eff}$, corresponding to the highest-$\dot{M}$ cases here. But at lower $\Sigma_{\rm eff}$ and $\dot{M}$, ignoring this term leads to order-of-magnitude or more over-estimation of $\dot{M}_{\rm acc}$, while incorporating the simple $1-f_{\rm wind}$ scaling predicted by Eq.~\ref{eqn:fwind} provides a remarkably good fit to the full simulation results (despite very different inflow structures in the different regimes; see \citealt{hopkins:zoom.sims}). 

Briefly, we note in applications of Eq.~\ref{eqn:fwind} in simulations which do explicitly include stellar feedback, that since BH accretion rates are generally evaluated in some resolution-scale kernel around the BH, one should evaluate and apply the $f_{\rm wind}$ correction within the approximately the same kernel, since that is precisely the scale where (by definition) explicit stellar feedback will cease to be resolved. But some care is needed and (like with any sub-grid model) the range of applicable scales is finite. If, for example, the unresolved region is so small that the radial infall timescale for the gas is much shorter than the timescale for stars to form and begin producing feedback there (e.g.\ $\lesssim 10^{5}$\,yr), then stellar feedback should not have a noticeable effect on SMBH accretion on these scales.

\section{Consequences}
\label{sec:consequences}

\begin{figure}
    \centering
    \includegraphics[width=0.96\columnwidth]{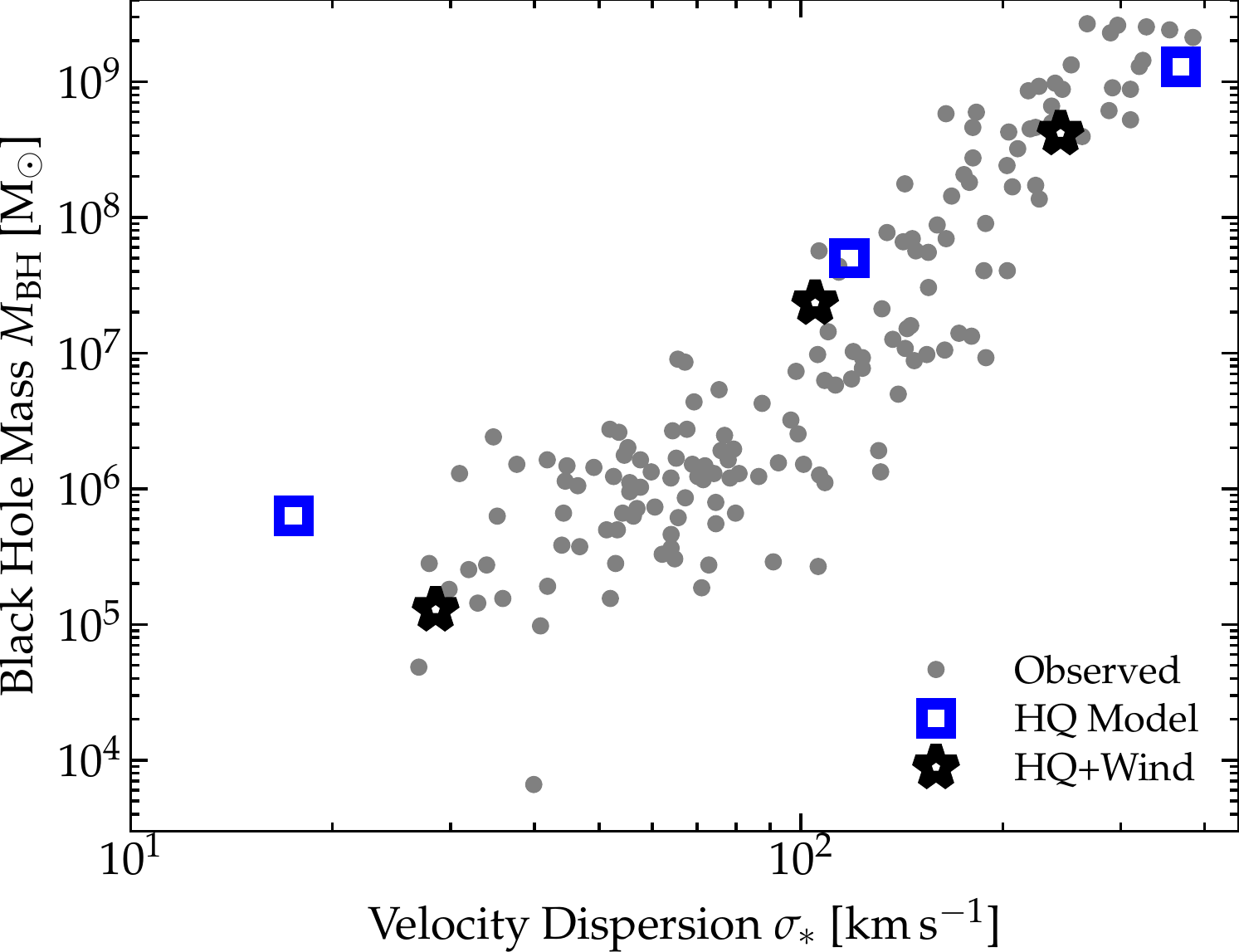}
    \vspace{-0.2cm}
    \caption{Preliminary comparison of cosmological simulations of three galaxies (to be studied in Wellons et al., in prep), with star formation and stellar feedback following the FIRE project \citet{hopkins:fire2.methods}, including accreting BHs as in e.g.\ \citet{daa:BHs.on.FIRE}, using two different sub-grid accretion models: (1) the HQ model as in Fig.~\ref{fig:inflow.sims} and  \citet{angles.alcazar:grav.torque.accretion.cosmo.sim.implications}, and (2) the identical prescription adding the same $(1-f_{\rm wind})$ correction factor from Fig.~\ref{fig:inflow.sims} and Eq.~\ref{eqn:fwind}. We compare to the compilation of observed BHs in \citet{baldassare:2020.low.mass.mbh.sigma}. For massive systems the proposed correction is a minor effect on total cosmic BH growth, but for dwarfs without massive bulges, ignoring the proposed correction could significantly over-predict BH growth.}
    \label{fig:m.sigma}
\end{figure}

This simple analytic expression has a number of interesting scaling properties and consequences. In a time-averaged sense, ignoring variations in accretion efficiency through the BH accretion disk, $\dot{M}_{\rm BH} \propto \eta \, (1-f_{\rm wind})\,M_{\rm gas}(<R)\,\Omega \sim  \eta \, (\Sigma_{\rm eff}/(\Sigma_{\rm crit} + \Sigma_{\rm eff}))\,M_{\rm gas}(<R)\,\Omega$. BH growth is dominated by episodes at high accretion rates, which for this estimator are dominated by periods with high gas fractions and $\Sigma_{\rm eff} \gtrsim \Sigma_{\rm crit}$ in the central $\sim$\,kpc; these have a characteristic integrated duration $\Delta t \sim {\rm a\ few}\ t_{\rm dyn} = \tau/\Omega$ (where $\tau\sim$ a few, before star formation, outflows, or accretion itself deplete the gas),\footnote{In our order-of-magnitude arguments here, it makes no difference whether the BH grows most of its mass in a single ``event'' with duration $\sim \Delta t$, or several events with similar conditions and total (sum) duration $\sim \Delta t$. If BHs grow primarily via many independent events which each contributes very little mass, or via BH-BH mergers, a different treatment would be needed.} so $M_{\rm BH} \sim \dot{M}_{\rm BH}\,\Delta t$. With this toy model in mind, consider:
\begin{itemize}

\item{\bf The Connection Between BHs and Bulges:} Because $\dot{M}_{\rm BH}$ decreases rapidly when $\Sigma_{\rm eff} \ll \Sigma_{\rm crit}$, the final BH mass is essentially proportional to the mass of gas at $\Sigma_{\rm eff} \gtrsim \Sigma_{\rm crit}$ (most of which forms stars, as $f_{\rm wind} < 1$ at these densities) in the galaxy center. In other words, the BH growth is specifically sensitive primarily {\em to the mass at high surface densities} in the galaxy center. But in nearly all studies of BH-host galaxy scalings, the ``bulge'' is {\em defined photometrically as excess light above the central surface brightness of the disk} \citep{KormendyRichstone95,kormendy99,magorrian,FM00,Gebhardt00}. This is in fact how such bulges (or NSCs) are usually observationally defined (and measured via e.g.\ B/D decomposition; \citealt{ferrarese:nuclear.cluster.vs.host.mass}). The critical surface mass density $\Sigma_{\rm crit} \sim 3\times10^{9}\,{\rm M_{\odot}\,kpc^{-2}}$ corresponds, for an old stellar population, to a B-band surface brightness $\mu_{\rm B}^{\rm crit} \sim 18-20\,{\rm mag\,arcsec^{-2}}$ (ignoring surface-brightness dimming at high redshifts). This corresponds very neatly with typical $\mu_{\rm B}$ above which bulges or NSCs appear \citep[references above and e.g.][]{allen:bulge-disk,fisher:pseudobulge.ns}! That is not an accident, as the {\em same} $\langle \dot{p}/m_{\ast} \rangle$ or $\Sigma_{\rm crit}$ appears (via $f_{\rm wind}$) in the self-regulation of star formation that regulates galaxy mass profiles/surface densities \citep[see][]{grudic:max.surface.density,grudic:mond.accel.scale.from.stellar.fb}. But the physical interpretation is quite different, as here it is not AGN but stellar feedback doing the ``regulation.''

What is striking here is that, unlike many BH accretion rate models, this depends {\em explicitly} on surface brightness/density (the same quantity that defines bulges/NSCs), in a {\em non-linear} manner. This provides an obvious, natural explanation for the fact that BHs appear to better correlate with the properties of these ``central light'' excesses, instead of just the galaxy properties as a whole, or the central potential (which, especially in disks with small bulges/NSCs, can easily be dominated by the more extended DM halo and disk), or properties of the disk, or circular velocity/halo mass \citep{tremaine:msigma,gultekin:msigma.update,kormendy:2011.bh.nodisk.corr,kormendy:2011.bh.nohalo.corr,reines:2015.dwarf.gal.mbh.mgal.norm.dift.and.huge.scatter}. It also naturally explains secondary correlations with Sersic index \citep{graham:sersic,graham:2015.mbh.host.steep.at.lowmass}, as higher $n_{s}$ is a direct reflection of the central high-$\Sigma$ light component, and why ``pseudobulges'' as defined in e.g.\ \citet{kormendy.kennicutt:pseudobulge.review,fisher:pseudobulge.ns,kormendy:2012.spheroidals}, which feature disk-like low $n_{s}$ (flat/low central surface brightness profiles) correlate more poorly with BH mass \citep{greene:pseudobulge.msigma,hu:msigma.pseudobulges,fisher:2012.mol.gas.vs.classical.pseudo.bulge,kormendy:2013.review.smbh.host.correlations}.

\item{\bf The $M_{\rm BH}-\sigma$ Relation:} The central velocity dispersion of a galaxy scales as $\sigma^{2} \sim G\,M_{\rm tot}(<R_{e})/R_{e}$. For galaxies (including most disk+bulge systems) where at the effective radius $R_{e}$, the effective surface density $\Sigma_{\rm eff}$ is below $\Sigma_{\rm crit}$ (i.e.\ $\mu_{\rm eff} \gtrsim \mu_{\rm crit}$, where $\mu_{\rm crit}$ and $\mu_{\rm crit}$ are the approximate surface brightness values in some band corresponding to stellar surface densities of $\sim\Sigma_{\rm eff}$ and $\sim \Sigma_{\rm crit}$, respectively), this implies $M_{\rm BH} \sim \tau\,\eta \, (\Sigma_{\rm eff} / \Sigma_{\rm crit})\, M_{\rm gas} \sim (\tau\,f_{\rm gas}/G^{2}\Sigma_{\rm crit})\,\eta\, (G\,M_{\rm tot}/R)^{2} \sim (\tau\,f_{\rm gas}\,\eta/G^{2}\Sigma_{\rm crit})\, \sigma_{\rm 3D}^{4} \sim 10^{8.5}\,M_{\odot}\,(\tau\,f_{\rm gas}\,\eta/0.001)\,(\sigma_{\rm 1D}/200\,{\rm km\,s^{-1}})^{4} \propto \sigma^{4}$, in excellent agreement with the relation observed \citep{gultekin:msigma.fits,kormendy:2013.review.smbh.host.correlations,2016ApJ...825....3L}, {\em especially} for low-mass BHs in small/dwarf/late-type host galaxies with effective surface densities $\ll \Sigma_{\rm crit}$ \citep{barth:2004.pox52.bh.host.props,peterson:2005.ngc4395.lowmass.bh.host,baldassare:2015.rg118.lowmbh.detected}. This is demonstrated explicitly in preliminary cosmological simulation tests in Fig.~\ref{fig:m.sigma}. Note that this is similar to the derivation in \citet{king:msigma.superfb.1,murray:momentum.winds,2006ApJ...650L..37M} of $M_{\rm BH} \propto \sigma^{4}$ for self-regulation via single-scattering radiation pressure (momentum flux $\dot{p} = L/c$) for an Eddington-limited BH, not by accident, because $\langle \dot{p}/m_{\ast}\rangle$ is order-of-magnitude similar to $\sim L/c$ for the stars (whether it comes in actual radiation, stellar winds, or SNe) and the $L$ of young stars is dominated by approximately Eddington-limited massive stars (see \citealt{grudic:mond.accel.scale.from.stellar.fb}).

\item{\bf The $M_{\rm BH}-M_{\rm bulge}$ Relation:} On the other hand, if most of the galaxy stellar mass lies above $\Sigma_{\rm crit}$ ($\mu_{\rm eff} \lesssim \mu_{\rm crit}$, i.e.\ ``pure (classical/dense) bulge'' systems), then $1-f_{\rm wind} \sim 1$, and the SFE is order-unity, so we simply have $M_{\rm BH} \sim \eta\,M_{\ast} \sim 0.001\,(\tau\,\eta/0.001)\,M_{\rm bulge}$. In other words, going from $\mu_{\rm eff} \gg \mu_{\rm crit}$ to $\mu_{\rm eff} \ll \mu_{\rm crit}$, this predicts a transition from $M_{\rm BH}-\sigma$ to $M_{\rm BH}-M_{\rm bulge}$ being the more ``causal'' or ``intrinsic'' relation. This is somewhat similar to suggestions of a ``break'' in $M_{\rm BH}-\sigma$ owing to the well-observed break in the Faber-Jackson relation, where dry-merging would lead to a dominant $M_{\rm BH}-M_{\rm bulge}$ relation at larger masses \citep{aller:mbh.esph,lauer:massive.bhs,mcconnell:mbh.host.revisions,graham:2015.mbh.host.steep.at.lowmass,2019ApJ...887...10S,2021A&A...649A.119P}, but in this case the discriminating criterion is surface brightness-based.\footnote{Of course, such a break owing to the role of dry-merging around $\sim 10^{10}\,{\rm M_{\odot}}$ would owe in part to galactic star formation being quenched at higher masses, which may relate to $\Sigma_{\rm crit}$ as discussed herein \citep{2021A&A...649A.119P}. This means these predictions may all be coupled in a non-trivial manner.} However it is important to note the caveat that there is no obvious difference in the scatter between $M_{\rm BH}(\sigma)$ and $M_{\rm BH}(M_{\rm bulge})$ observed at present in massive ellipticals \citep{mcconnell:mbh.host.revisions,2019ApJ...876..155S}. But again, because of the natural connection to surface density/acceleration, this argument would explain why the lowest-mass BHs in small hosts (with photometric ``bulges'' with relatively low central surface brightness) appear to be ``low'' relative to an extrapolated $M_{\rm BH}-M_{\ast}$ relation while agreeing better with $M_{\rm BH}-\sigma$ (see Fig.~\ref{fig:m.sigma}  and \citealt{barth:2004.pox52.bh.host.props,peterson:2005.ngc4395.lowmass.bh.host,green:low.m.bhs,kormendy:2013.review.smbh.host.correlations,baldassare:2015.rg118.lowmbh.detected}).

\item{\bf Mild Redshift Evolution:} It is well-established that the progenitors of giant elliptical galaxies today had their central, high-surface brightness ``cores'' in place at high redshifts $z\gtrsim 2$ \citep{hopkins:density.galcores,bezanson:massive.gal.cores.evol}, and grew primarily in both size and mass via dry merging of smaller systems which accrete the extended ``envelope'' of low-surface brightness material and ICL \citep{van-dokkum:compact.e.evol.w.profile.changes,wellons:2016.evolutionary.paths.massive.galaxies}. These ``cores'' (whether cuspy or ``cored'' in their nuclear profile) easily exceed $\Sigma_{\rm crit}$; so if the BH is sensitive to the mass above $\Sigma_{\rm crit}$ it would reflect essentially the entire galaxy mass in the progenitor. The subsequent merging would contribute negligible material at $>\Sigma_{\rm crit}$, and even the merging BHs are unlikely to sink via dynamical friction \citep{hopkins:groups.ell,hopkins:cusps.evol,hopkins:cores}, so $M_{\rm bulge}$ will increase but $M_{\rm BH}$ will not, leading to redshift evolution in $M_{\rm BH}/M_{\rm bulge}$ \citep[as proposed in][]{croton:msigma.evolution,hopkins:r.z.evol}. But the effect would be mild, because these galaxies have probably only grown by a factor of $\sim 2$ in stellar mass (making this the upper limit to redshift evolution in $M_{\rm BH}/M_{\rm bulge}$ to $z\sim2-4$), consistent with observational limits \citep{suh:2020.no.bh.host.gal.evol.scaling}.

\item{\bf ``Outliers'' in Compact Es:} For the same reasons, at similar total $M_{\rm bulge}$, high-surface brightness cEs will have most of their stellar mass at densities $>\Sigma_{\rm crit}$, while giant Es might have a significant mass fraction below $\Sigma_{\rm crit}$, implying the cE would have a larger $M_{\rm BH}$ from these scalings. This is consistent with some claims for observed ``outliers'' \citep{mcconnell:mbh.host.revisions,seth:ultra.massive.bh.in.ucd,2015Sci...349..168T,2016ApJ...817....2W,2020arXiv200108753L}; however, we stress that the effect saturates, as once most of the mass is at $\Sigma_{\rm eff} \gtrsim \Sigma_{\rm crit}$, there is no ``additional'' dependence on compactness, also consistent with the relatively modest limits on such dependence in e.g.\ \citet{ni:2019.bh.vs.gal.compactness}.

\item{\bf Quenching and Central Surface Densities (``$\Sigma_{1}$''):} In the last few years studies have shown that a number of galaxy and BH properties, particularly related to ``quenching,'' are closely correlated with the central surface density of the galaxy, often parameterized as ``$\Sigma_{1} \equiv M_{\ast}(<1\,{\rm kpc})/\pi\,({\rm 1\,kpc})^{2}$ \citep{franx:2008.surface.density.dependent.quenching.evolution,cheung:2012.sigma.1.central.density.vs.quenching,vanderwel:2012.candels.structural.parameters,whitaker:2012.sf.vs.mass.density.vs.redshift,huertas.company:2016.surface.density.vs.quenching.morph,ellison:2018.sf.profiles.surface.densities.quenching,lee:2018.central.density.vs.galaxy.properties}. It is immediately obvious that the model here predicts such a correlation with BH growth: for example, this would automatically explain recent studies showing that BH growth rates and AGN activity increases with $\Sigma_{1}$ at otherwise fixed galaxy properties \citep{ni:2019.agn.activity.bh.growth.sigma.1.galaxy.compactness}. But more strikingly, the most robust observation of interest is that $\Sigma_{1}$ correlates strongly with whether or not a galaxy is ``quenched,'' with the quenched fraction increasing rapidly around a critical $\Sigma_{1} \sim 3\times10^{9}\,M_{\odot}\,{\rm kpc^{-2}}$ \citep{cheung:2012.sigma.1.central.density.vs.quenching,barro:2017.critical.sigma.1.ridgeline.for.quenching} -- remarkably similar to the predicted $\Sigma_{\rm crit}$! Phenomenologically, many have argued this could be a signature of quenching driven by AGN feedback, if BHs were somehow sensitive to $\Sigma_{1}$ \citep{pandya:2017.transition.galaxies.properties.vs.sims,rodriguez.puebla:2017.quenching.vs.sigma.1.bh.mass,chen:2020.overmassive.bh.quenching.model}. The models here predict a natural explanation for precisely such a dependence of BH growth, and therefore AGN feedback, on $\Sigma_{1}$ around $\Sigma_{\rm crit}$: for example, in a model where e.g.\ the integrated BH feedback energy deposition scales $E_{\rm fb} \sim \int \epsilon_{\rm fb}\,L_{\rm AGN} \,dt \sim \epsilon_{\rm fb}\,0.1\,M_{\rm BH}\,c^{2} \sim \epsilon_{\rm fb}\,0.1\,\eta\,(\Sigma_{1}/\Sigma_{\rm crit})\,M_{\rm gas}$ (using the scalings above), comparing this to the binding energy of the halo gas ($E_{\rm halo} \sim f_{\rm bar}\,V_{c}^{2}\,M_{\rm halo}$) assuming a universal baryon fraction $M_{\rm gas} \sim f_{\rm bar}\,M_{\rm halo}$, we have $E_{\rm fb} \gtrsim E_{\rm halo}$ for $\Sigma_{1} \gtrsim \Sigma_{\rm crit}\,(0.01/\epsilon_{\rm fb})\,(0.001/\eta)\,(V_{c}/200\,{\rm km\,s^{-1}})^{2}$, remarkably similar to the observed quenching ``ridgeline'' \citep{chen:2020.overmassive.bh.quenching.model}.

\item{\bf The $L_{\rm AGN}$-SFR Relation:} It is observationally well-established that galactic star formation scales with surface density \citep{kennicutt98}. In fact, standard theoretical models of the Kennicutt-Schmidt (KS) relation generically predict $\dot{\Sigma}_{\ast} \sim t_{\ast}^{-1}\,(\Sigma_{\rm eff}/\Sigma_{\rm crit})\,\Sigma_{\rm gas}$,\footnote{Here $t_{\ast} \equiv \langle p/{m}_{\ast} \rangle/\langle\dot{p}/{m}_{\ast} \rangle = (\int \langle \dot{p}[t]/\dot{m}_{\ast} \rangle\,dt) / \langle \dot{p}[t=0]/\dot{m}_{\ast} \rangle \sim 100\,{\rm Myr} $ where $ \langle p/{m}_{\ast} \rangle$ is the time-integrated momentum injected by a single stellar population (SSP) while $\langle \dot{p}/m_{\ast} \rangle$ is the instantaneous rate for a zero-age SSP (both averaged over the stellar IMF).} with the same $\Sigma_{\rm crit}$ appearing because stellar feedback self-regulates the local SFR \citep{ostriker.shetty:2011.turb.disk.selfreg.ks,hopkins:rad.pressure.sf.fb,cafg:sf.fb.reg.kslaw}. Combining this with our expression for $\Sigma_{\rm eff} \lesssim \Sigma_{\rm crit}$ (where these derivations of the KS relation are valid), we immediately obtain $\langle\dot{M}_{\rm acc}\rangle \sim  \eta\,(t_{\ast}\,\Omega)\,\dot{M}_{\ast}$ where $t_{\ast}\,\Omega\sim 1$ is only weakly dependent on galaxy properties. Using standard bolometric conversions this can be written as $\langle L_{\rm AGN} \rangle \sim 0.1\,(\eta\,t_{\ast}\,\Omega/0.001)\,\langle L_{\rm SF} \rangle$ (or e.g.\ X-ray AGN luminosity vs. IR luminosity from star formation: {$L_{\rm X,\,AGN} \sim 0.004\,(\eta\,t_{\ast}\,\Omega/0.001)\,L_{\rm IR}$}), in excellent agreement with the observed relation when AGN variability and selection effects are properly included \citep{hickox:variability.and.linear.sfr.mdot.corr,grimmett:2020.agn.l.vs.sfr.continuous.notbinned}. Indeed, observations may specifically indicate a closer correlation between BH accretion and bulge/compact star formation (as compared to galaxy-wide SFRs), which would naturally follow from this \citep{2019MNRAS.485.3721Y,2021MNRAS.500.4989N}.

\item{\bf Off-Nuclear Fueling/AGN:} Although accretion models of the form in Eq.~\ref{eqn:mdot.basic} do not require (unlike e.g. Bondi-Hoyle accretion) that the BH dominates the potential on all scales, they {\em do} assume that the BH resides near the local center/minimum of the potential, so that gas which loses angular momentum or energy tends (on average) to move ``inwards'' or ``towards'' the BH. If a BH is ejected or free-moving through the galaxy (as seen in many dwarfs; \citealt{reines:2020.off.nuclear.agn.in.dwarfs}), this is no longer valid and Eq.~\ref{eqn:mdot.basic} will tend to over-estimate $\dot{M}_{\rm BH}$. While our simple $(1-f_{\rm wind})$ correction is not specifically designed to address this situation, it does have the effect of reducing $\dot{M}_{\rm BH}$ when BHs are off-center, as $\Sigma_{\rm eff}$ is lower, providing at least a partial improvement in accuracy.

\end{itemize}

\section{Conclusions}
\label{sec:conclusions}

We consider the simplest-possible extension to standard models of AGN/SMBH accretion (parameterized as $\dot{M}_{\rm acc} \sim \eta(...)\,M_{\rm gas}(<r)\,\Omega(r)$) from galactic nuclei scales ($\sim 0.1-1000\,$pc), to account for the role of {\em stellar} feedback ejecting gas from smaller scales before it reaches the AGN accretion disk. As shown in \citet{torrey.2016:fire.galactic.nuclei.star.formation.instability}, when the dynamical time $t_{\rm dyn} \sim \Omega^{-1} \sim r/V_{c}$ is less than the stellar evolution timescale for most SNe ($t_{\ast} \sim 100\,$Myr), the presence of gas in galactic nuclei (and hence its ability to accrete further inwards) is regulated by stellar feedback, with efficient feedback able to eject most gas from the nucleus (not necessarily the galaxy) at low densities. Simple analytic models, detailed simulations of molecular clouds and nuclear gas disks, and direct observations all argue that the efficiency of this ejection scales in a simple manner with the gravitational acceleration $\bar{a}_{\rm grav}\equiv G\,M_{\rm enc}(<r)/r^{2}$ or ``effective surface density'' $\Sigma_{\rm eff} \equiv M_{\rm enc}(<r)/\pi\,r^{2}$, as $M_{\rm ejected}/M_{\rm retained} \sim \langle \dot{p}/m_{\ast} \rangle / \bar{a}_{\rm grav} \sim \Sigma_{\rm crit} / \Sigma_{\rm eff}$, where $\langle \dot{p}/m_{\ast} \rangle \sim 10^{-7}\,{\rm cm\,s^{-2}}$ ($\Sigma_{\rm crit} \equiv \langle \dot{p}/m_{\ast} \rangle / \pi\,G \sim 3\times10^{9}\,{\rm M_{\odot}\,kpc^{-2}}$) is the momentum flux per unit mass in feedback (radiation+stellar mass-loss+SNe) from a zero-age main sequence IMF-integrated stellar population. This leads to a ``correction factor'' to accretion models which ignored such stellar feedback-driven ejection, of the form $\eta \rightarrow \eta\,(1-f_{\rm ejected}) \sim \eta\,\Sigma_{\rm eff}/(\Sigma_{\rm eff} + \Sigma_{\rm crit})$. 

We show that this immediately resolves some discrepancies between various high-resolution simulation studies of accretion and inflows in galactic nuclei. Simulations which included explicit resolved stellar feedback, but focused on quasar-level, extremely dense gaseous torii or nuclear disks with $\Sigma_{\rm eff} \gtrsim 10^{5}\,{\rm M_{\odot}\,cm^{-2}} \gg \Sigma_{\rm crit}$ \citep{hopkins:qso.stellar.fb.together,wada:torus.mol.gas.hydro.sims} have found accretion rates $\dot{M}_{\rm acc}$ in good agreement with older simulations that did not include explicit stellar feedback-driven outflows at all \citep[e.g.][]{hopkins:zoom.sims,hopkins:cusp.slopes,hopkins:inflow.analytics}, while simulations with lower central densities (representing disks or dwarf galaxies, with little nuclear gas) found much lower inflow rates \citep{dubois:delayed.cooling.sne.models,torrey.2016:fire.galactic.nuclei.star.formation.instability,daa:BHs.on.FIRE}. 

We go on to show that with this correction factor, the resulting approximate expression for BH accretion rates has a number of interesting properties. Most importantly, because $1-f_{\rm ejected} \sim 1$ when $\Sigma_{\rm eff} \gtrsim \Sigma_{\rm crit}$ and $1-f_{\rm ejected} \sim \Sigma_{\rm eff} / \Sigma_{\rm crit} \ll 1 $ when $\Sigma_{\rm eff} \ll \Sigma_{\rm crit}$, this predicts that BH mass should be correlated most directly with the mass in the galaxy center above a critical effective (total-mass) surface density $\sim \Sigma_{\rm crit}$ (which, from the same feedback-regulation model, should mostly turn into stars at these high densities). This corresponds to an intrinsic stellar surface brightness $\mu_{\rm B} \sim (18-20)\,{\rm mag\,arcsec^{-2}}$ for an old stellar population (depending on age, metallicity, etc). This corresponds remarkably well to the characteristic surface brightness above which ``bulges'' dominate the light. In fact, in essentially all studies of BH-host galaxy scalings, ``bulge'' properties are defined {\em photometrically}, as excess surface brightness above the disk around the BH. This is true even when the bulge does not contain enough total mass to dominate the central potential or escape velocity from the galaxy center -- where models which predict BH mass traces binding/kinetic/potential energy or escape velocity would predict a better correlation between BH properties and disks, instead of bulges (which is not observed). This also immediately explains why BHs do not simply correlate with ``central mass'' within some fixed physical aperture, as many models also predict, but with specific photometric {\em features} of galaxies. In short, this simple stellar-feedback-regulated scaling therefore immediately explains why, in fact, {\em bulge} properties appear to predict BH masses. 

We also show that this scaling leads immediately to the observed BH-$\sigma$ relation, directly, especially in lower-mass host galaxies, and explains a wide variety of secondary correlations or lack thereof (e.g.\ why BHs appear to correlate more poorly with photometrically defined ``pseudobulges''; secondary correlations with galaxy compactness, Sersic index, redshift, and position on the Faber-Jackson relation). 
And we show that, during active accretion phases, if we invoke the same stellar-feedback regulated arguments commonly used to explain the galactic Schmidt-Kennicutt star formation scalings, we immediately predict a correlation between {\em mean} AGN luminosity (albeit with large variability expected) and galactic SFR, in agreement with that observed.
As a result, if AGN feedback plays a critical role in galaxy quenching, the argument here may also play a critical role explaining the ``critical'' value of central surface density $\Sigma_{1} \sim 3\times 10^{9}\,{\rm M_{\odot}\,kpc^{-2}}$ above which galaxies tend to be quenched -- which is observed to be remarkably  similar to the predicted $\Sigma_{\rm crit}$ where AGN accretion is efficient.

Of course, our study here is a simple analytic investigation of dimensional scalings. More refined models will require further, high-resolution numerical simulations and observations of gas in galactic nuclei to test these scalings and calibrate exact coefficients as well as detailed dependence on e.g.\ gas properties, stellar mass distributions, dynamical state of galaxies, etc. We also stress that we neglect AGN feedback here, as an additional regulator of BH accretion. Of course, AGN can eject mass directly from accretion disk/jet scales (this would appear as some sub-grid ``efficiency'' in models here); they can also regulate inflow on these scales by driving large-scale outflows, changing the properties (e.g.\ $M_{\rm gas}$, $\Sigma_{\rm eff}$) which determined $\dot{M}_{\rm acc}$, but this does not necessarily change our scaling (Eq.~\ref{eqn:mdot.basic}) for $\dot{M}_{\rm acc}$. Determining whether there is a more complex non-linear interplay again requires self-consistent simulations.

\acknowledgments{We thank Jessie Christiansen, Michael Fall, Peter Mitchell, Guang Yang, and our anonymous referee for a number of helpful suggestions and insightful comments. Support for PFH was provided by NSF Research Grants 1911233 \&\ 20009234, NSF CAREER grant 1455342, NASA grants 80NSSC18K0562, HST-AR-15800.001-A. 
SW is supported by an NSF Astronomy and Astrophysics Postdoctoral Fellowship under award AST2001905. CAFG was supported by NSF through grants AST-1715216 and CAREER award AST-1652522; by NASA through grant 17-ATP17-0067; by STScI through grant HST-AR-16124.001-A; and by the Research Corporation for Science Advancement through a Cottrell Scholar Award and a Scialog Award. 
Numerical calculations were run on the Caltech compute cluster ``Wheeler,'' allocations FTA-Hopkins supported by the NSF and TACC, and NASA HEC SMD-16-7592.}

\datastatement{The data supporting the plots within this article are available on reasonable request to the corresponding author. A public version of the GIZMO code is available at \gizmourl.}

\bibliography{ms_extracted}

\end{document}